\documentclass[prl,amssymb,amsfonts,twocolumn]{revtex4}
\usepackage{amsmath,amsthm,amsfonts,graphicx,epsfig}
\usepackage[usenames]{color}
\newcommand{\ket}[1]{\left| #1 \right\rangle}
\newcommand{\bra}[1]{\left\langle  #1 \right|}
\newcommand{\braket}[2]{\langle #1 | #2 \rangle}

\newcommand{\dbar}{\kern-.1em{\raise.8ex\hbox{ -}}\kern-.6em{d}}

\def\half{\mbox{$1\over2$}}

\def \be{\begin{equation}}
\def \ee{\end{equation}}
\def \barray{\begin{eqnarray}}
\def \earray{\end{eqnarray}}
\begin{document}
\title{Entanglement on demand through time reordering}
\author{J. E. Avron, G. Bisker,  D. Gershoni, N. H.
Lindner, E. A. Meirom} \affiliation{Department of Physics,
Technion---Israel Institute of Technology, 32000 Haifa, Israel}
\author{R. J. Warburton}
\affiliation{School of Engineering and Physical Sciences,
Heriot-Watt University, Edinburgh EH14 4AS, United Kingdom}

%\date{\today}%
%\maketitle
\begin{abstract}

Entangled photons can be generated ``on demand'' in a novel scheme
involving unitary time reordering of the photons emitted in a
radiative decay cascade. The scheme yields polarization entangled photon
pairs, even though prior to reordering the emitted photons carry
significant ``which path information'' and their polarizations are
unentangled. This shows that quantum chronology can be manipulated
in a way that is lossless and deterministic (unitary). The theory
can, in principle, be tested and applied to the biexciton cascade
in semiconductor quantum dots.

\end{abstract}

\pacs {03.67.Mn, 78.67.Hc,  42.50.Dv, 03.65.Ud, 03.67.Bg}

\maketitle

%%%%%%%%%%%%%%%%%%%%%%%%%%%%%%%%%%%%%%%%%%%%%%%%%%%%%%%%%%%%%%%%%%%%
%%%%%%%%%%%%%%%%%%%%%%%%%%%%%%%%%%%%%%%%%%%%%%%%%%%%%%%%%%%%%%%%%%%%
%%%%%%%%%%%%%%%%%%%%%%%%%%%%%%%%%%%%%%%%%%%%%%%%%%%%%%%%%%%%%%%%%%%%
Entangled quantum states are an important resource in quantum
information and communication \cite{Nielsen-Chuang}. Entangled
photons are particularly attractive for applications due to their
non interacting nature and the ease by which they can be
manipulated. There is, therefore, considerable interest in the
development of sources for reliable (non-random) polarization
entangled photon pairs. Currently, the most important and
practical source of polarization entangled pairs is down
conversion
\cite{down-conversion-Bouwmeester,down-conversion-Kwiat} which has
a large random component. Furthermore, since the entanglement is
created by a coincidence detection of the pair, the entangled
state becomes unavailable for further manipulations.

\begin{figure}[ht]
%\begin{left}
%  \includegraphics[width=10cm]{system_setup_new.eps}
\epsfxsize=.34\textwidth\centerline{\epsffile{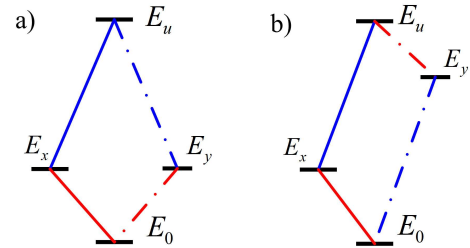}}
  \caption{Two alternative level schemes that can be used to
  generate entangled pairs. The solid (dashed-dotted) line represents
  the decay channel that yields two $x$ ($y$) polarized photons.
  The energies correspond to the colors of the emitted photons.
  (a) Represents the situation where photon colors match in the first generation,
  and has the geometry of a kite (deltoid);
  (b) Represents the case where colors in different generation match,
   and is geometrically a parallelogram.
}
  \label{fig:2decaycascades}
%\end{left}
\end{figure}

Quantum dots are a source of single photons ``on demand''
\cite{Michler,Santori,Yuan}. Recently, it has been shown
\cite{akopian,shields} that they can be used as sources of
polarization entangled photon pairs. The entangled photon pair is
obtained from the decay cascade of a biexciton in a quantum dot. The
biexciton has two decay channels, each emitting a photon pair with a
polarization characteristic to the channel. Perfect ``which path
ambiguity'' requires that the intermediate exciton state is doubly
degenerate as illustrated in Fig.~\ref{fig:2decaycascades} (a).
In this idealized setting, the first generation photons have
identical colors (energies) and the second generation photons also have identical, though in general different, colors. With perfect
``which path ambiguity'' the state of polarization of the two
photons is maximally entangled, and each pair can, in principle, be
produced ``on demand'' \cite{Benson}.

Quantum dots do not have perfect cylindrical symmetry  and this
lifts the degeneracy of the intermediate exciton states
\cite{bayer}. We shall refer to this splitting as ``detuning''.
Since the detuning is large (compared with the radiative width),
the two decay paths are effectively distinguished by the distinct
colors of the emitted photons. This adversely affects the ``raw''
entanglement which is then negligible \cite{akopian}.

In principle, the detuning can be manipulated by Stark and Zeeman
effects, by stress etc. and much experimental effort has been
devoted into reducing it to small values (below the radiative
width) \cite{ref:experiments-degeneracy}. This has been an elusive
goal so far for both practical and fundamental reasons. The
fundamental reason is that quantum mechanics has the principle of
level repulsion: In quantum dots the scale of energy responsible
for the detuning (exchange) dominates the scale of the radiative
width which is the smallest energy scale in the problem
\cite{bayer,bimberg}. This puts an ``in principle'' obstacle to
substantial ``which path ambiguity'' in quantum dots.

Entangled photons from quantum dots have been obtained
by selectively filtering the photons that conform to the which
path ambiguity \cite{akopian}. The entanglement then comes at the
price that a substantial fraction of the photon pairs are lost and
the quantum dot does not furnish entangled pairs ``on demand''.

An alternate strategy proposed in \cite{finley,Reimer} is to tune
the dot spectrum to have coincidence of colors {\em across}
generations, rather than {\em within} generations. This is
illustrated in Fig.~\ref{fig:2decaycascades}~(b) where the colors
of the photons in the first generation match, in pairs, the colors
of the photons in the second generation. Since a coincidence of
colors in different generations does not require degeneracy, there
is no {\em  fundamental} obstacle to tuning the level diagram to a
precise coincidence of colors \cite{bimberg,finley,Reimer}. When
this is the case the two decay channels are identical up to time
ordering. The different time ordering of the two decay paths of
the `raw' state betray the path which, as it turns out, completely
kills the entanglement.
%It was
%suggested in \cite{finley} that using methods of linear optics one
%might be able to reorder the photons and regain entanglement.%
%
%
%
In order to regain entanglement one needs to manipulate the time
ordering.
%Although the idea is appealing, it does not address conceptual
%and practical issues and it calls for the development of a theory of
%re-ordering of quantum chronology.
The theory of re-ordering the quantum chronology is developed in
this Letter. It allows us to derive the measure of entanglement of
the reordered state and its dependence on the spectral properties
of the radiative cascade. Perhaps the most important result is
that it shows that the reordering can be made in a way that
conforms with the requirement of ``on demand''.
%(in contrast with what is stated in \cite{Reimer}).
The theory then also allows an optimization of the entanglement and it
leads to a proposal of a practical experimental realization.

We denote by $\Delta$ the detuning, the (dimensionless) measure of
the color matching of the photons in a given generation. Perfect
matching within generation is represented by $\Delta=0$.
Similarly, we denote by $\beta$ the (dimensionless) spectral
control which measures the matching of the colors across
generations (the ``biexciton binding energy'' in a quantum dot).
Perfect cross color matching in this case is represented by
$\beta=0$. More precisely:
\begin{equation}
\Delta=\frac{E_y-E_x}{2\Gamma}, \quad
\beta=\frac{E_u-E_0-E_x-E_y}{2\Gamma}.
\end{equation}
$E_x,\ E_y,\ E_u$ are as in Fig.~\ref{fig:2decaycascades} and
$\Gamma$ is the half width of the intermediate levels. As we have
explained, in dots there are fundamental reasons that force
$|\Delta| \ge 1$, while $\beta$ can, in principle, be tuned to
zero. The issue is: Can one generate entangled photon pairs by
tuning $\beta=0$ even when $|\Delta|\gg 1$? As we show the answer
is yes.

%We denote by $\Delta$ the detuning, the (dimensionless) measure of
%the color matching of the photons in a given generation so that
%$\Delta=0$ represents perfect matching. Similarly, we denote by
%$\beta$ the (dimensionless) spectral control which measures the
%matching of the colors across generations (the ``biexciton binding
%energy'') so that, again $\beta=0$ represents perfect cross color
%matching. More precisely:
%\begin{equation}
%\Delta=\frac{E_y-E_x}{2\Gamma}, \quad
%\beta=\frac{E_u-E_x-E_y}{2\Gamma}
%\end{equation}
%$E_x,\ E_y,\ E_u$ are as in Fig.~\ref{fig:2decaycascades} and
%$\Gamma$ is the half width of the intermediate levels. $E_0$ is
%taken to be $0$. As we have explained, there are fundamental
%reasons that force $|\Delta| \ge 1$, while $\beta$ can, in
%principle, be tuned to zero. The issue is: Can one generate
%entangled photon pairs by tuning $\beta=0$ even when $|\Delta|\gg
%1$? As we shall see the answer is yes.

%%%%%%%%%%%%%%%%%%%%%%%%%%%%%%%%%
%%%%%%%%%%%%%%%%%%%%%%%%%%%%%%%%%%%%%%%%%%%%%
Suppose that the two photons are emitted along the z-axis and have
two decay modes with equal amplitudes. Then, the (possibly
un-normalized) photon wave function has the form
\cite{Cohen-Tannoudji}:
\begin{equation}\label{wave function of the emitted photons}
\left| \psi  \right\rangle  = \sum\limits_{j = x,y} \ket{\alpha_j}
\otimes \ket{jj}
\end{equation}
where $\ket{\alpha_j}$ describes the photons' wave packet and
$\ket{jj}$ their state of polarization.

For reasons that will become clear below we need to allow for a
unitary post processing of the `raw' state emerging from the
cascade. The manipulation, $U_j$, is described by a unitary
operation that depends on the polarization state (i.e. the decay
channel) and is then formally of the form
    \be
    \ket{\alpha_j}\to U_j\ket{\alpha_j}.
    \ee
In the language of quantum information this corresponds to
applying single qubit unitary gates on each of the two
polarization states (this operation can be made by whomever
prepares the state, but can also be made later and so falls under
the class of local operations \cite{horodecki-review}).

As a measure of the entanglement we take the absolute value of the
negative eigenvalue in the Peres test (negativity)
\cite{peres_test,entanglement_measure}
    \begin{equation}\label{gamma}
    \gamma(\Delta,\beta;W)=
    \frac{|\bra{\alpha_x}W\ket{\alpha_y}|}
    {\braket{\alpha_x}{\alpha_x}+\braket{\alpha_y}{\alpha_y}}, \quad
    W=U_x^*U_y.
    \end{equation}
The maximal value of $\gamma$ is $\half$ corresponding to
maximally entangled (Bell) states. Note that the denominator is
just the normalization of the state $\ket{\psi}$.

%As we shall see, maximal entanglement can be archived for
%$\beta=0$ and $\Delta \gg 1$ for an appropriate choice of $W$.

The $\ket{\alpha_j}$ are fully determined by the (complex)
energies of the level diagram, $Z_a=E_a-i\Gamma_a$. In the limit
that the dipole approximation holds, the (normalized) wave packets
are given by \cite{Cohen-Tannoudji,meirom}:
\begin{equation}\label{alphaj}
    \braket{k_1,k_2}{\alpha_j}=A\left(|k_1|+|k_2|,Z_u\right)
    \big(A\left(|k_1|,Z_j\right)+A\left(|k_2|,Z_j\right)\big)
\end{equation}
where:
\begin{equation}
\label{amplitude}
    A(k,Z)=\sqrt{\frac{\Gamma}{\pi}}\ \frac{1}{k-Z}\ .
\end{equation}
(We use units where $\hbar=c=1$.) The photon of the first
generation has energy near $E_u-E_j$ while the photon of the
second generation has energy near $E_j$ and relative time delay of
order $1/\Gamma_j$. Note that positive (negative) delay is
associated with $Z_j$ in the lower (upper) half complex
energy-plane.

In practice, the smallest energy scale in the problem is the
radiative width, $\Gamma$. We treat it as a small parameter in the
theory and thus can safely drop the absolute value in
Eq.~(\ref{alphaj}). This allows for an analytic calculation of some
of the integrals that arise. In particular, when the two emitted
photons have different colors, i.e. when
%$|\Delta|\gg 1$,
$||k_1|-|k_2|| \gg \Gamma$, one has
    \be\label{normalization}
    \braket{\alpha_j}{\alpha_j}=2%1
    \ee
to leading order in $\Gamma$.

The numerator in Eq.~(\ref{gamma}),
$\bra{\alpha_x}W\ket{\alpha_y}$ can now be written as a sum of the
two integrals:
 \begin{equation}\label{non direct}
    y_1= \frac{2\Gamma\Gamma_u}{\pi^2}\int
    \frac {W(k_1,k_2)\,dk_1 dk_2} {|k_1+k_2-Z_u|^2(k_1-Z_x^*)(k_2-Z_y)}
    \end{equation}
and
 \begin{equation}\label{direct}
    y_2=\frac{2\Gamma\Gamma_u}{\pi^2} \int   \
     \,\frac {W(k_1,k_2)\,dk_1  dk_2} {|k_1+k_2-Z_u|^2(k_1-Z_x^*)(k_1-Z_y)}
    \end{equation}
where $\Gamma_x=\Gamma_y = \Gamma$. The first term, $y_1$, may be
thought of as the contribution to the entanglement from the
coincidence of colors across generations while the second term,
$y_2$, as the contribution to entanglement from the matching of
colors within a generation.

Consider first the `raw' outgoing state where $W=1$. One expects
that entanglement can arise only from matching colors within a
generation. By first shifting the integration variables, $k_1- E_x
\to k_1$ and $k_2- E_y \to k_2$ in Eq.~(\ref{non direct}), and
using the residue theorem to compute the integrals, one indeed
finds that the cross-generations contribution vanishes:
    \begin{equation}
    \label{zero}
    y_1( \beta;W= 1)=0.
    %\!\!\frac{2\Gamma}{\pi} \int
   % \frac {dk_1  dk_2} {|k_1+k_2-2\Gamma \beta +i\Gamma_u|^2(k_1-i\Gamma)(k_2+i\Gamma)}
   %\frac {dk} {(-k-i(\Gamma+\Gamma_u)+2\Gamma \beta)(k+i\Gamma)}
   % \nonumber \\
    %&=&-\frac{2}{\pi} \int  \
    %\frac {dk} {(k-i)(k- 2\beta-i)}\nonumber \\
    %&=& 0
    \end{equation}
For the second integral, shifting $k_j\to k_j+\frac{E_x+E_y}2$
first, and using similar elementary manipulations, one finds
 \be
    y_2(\Delta;W=1)=
    %&=&\frac{2\Gamma}{\pi} \int
    %\frac {dk }
    %{(k+\Gamma\Delta -i\Gamma)(k-\Gamma\Delta+i\Gamma)}
    %\nonumber \\
    %    &=&\frac{2}{\pi} \int  \
    %   \frac {dk} {(k+\Delta -i)(k-\Delta+i)}\nonumber \\
    %   &=&
    \frac {-2i} {\Delta -i}.
    \ee
In particular, combining with
Eqs.~(\ref{gamma},\ref{normalization}), we find
\begin{equation}
\gamma^2(\Delta, \beta;1)=\frac {1} {4(\Delta^2 +1)}.
\end{equation}
The `raw' entanglement is {\em independent} of $\beta$ (to leading
order in $\Gamma$) and does not benefit from matching the colors of
photons in different generations (tuning to $\beta=0$). Since
typically $|\Delta| \gg 1$, the `raw' entanglement is small.

\begin{figure}[ht]
\begin{center}
   \epsfxsize=.5\textwidth \centerline{\epsffile{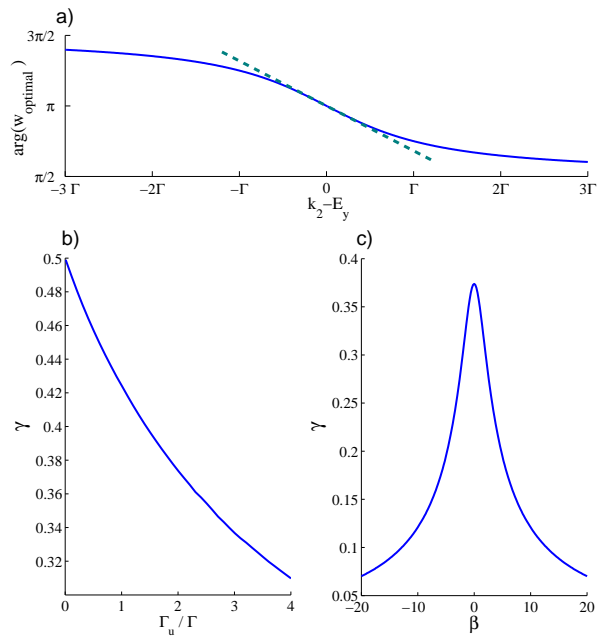}}
  \caption{ a) The argument of $W_{opt}$ as a function of $k_2-E_y$ for $k_1=E_x$.
  The dashed line represents a feasible linear
  approximation, generated by an optical delay.
  b) The off diagonal matrix element $\gamma$ for $W_{opt}$,
  as a function of the ratio $\Gamma_u/\Gamma$. For typical biexciton decays,
$\Gamma_u/\Gamma\approx 2$.
       c) $\gamma$ as a function of the color matching dimensionless parameter,
       $\beta$, for $\Gamma_u/\Gamma= 2$. The maximal value occurs when $\beta=0$.
       The width of the peak is of order $\Gamma$.} %\RED{GB}
  \label{fig:B_plot}
\end{center}
 \vspace{-0.3cm}
\end{figure}

Fortunately, a suitable choice of $W$ can yield an entangled state
of polarization even when $|\Delta| \gg 1$. Since $W$ is unitary,
$W(k_1,k_2)$ is a phase. The optimal choice of phase is one that
would make $y_1$ as large as possible. This can be achieved when the
integrand in Eq.~(\ref{non direct}) has a fixed phase (so that the
oscillations leading to the cancellation in Eq.~(\ref{zero}) are
eliminated), e.g.:
    \be\label{optimal}
    W_{opt}= - \frac {(k_1-Z_x^*)(k_2-Z_y)}
    {|k_1-Z_x^*||k_2-Z_y|}.
    \ee
$W_{opt}(E_x,k_2-E_y)$ is plotted in Fig.~\ref{fig:B_plot} (a).
%The physical meaning of the form of $W_{\rm opt}$ is explained
%later in the text.
%\REM{GB:order}

%Note that $W_{\rm opt}$ is of the form suggested by
%Eq.~\eqref{form W}. %
By inserting Eq.~(\ref{optimal}) into Eq.~(\ref{non direct}) and
shifting the integration variables, this choice for $W_{opt}$
gives for $y_1$
 \barray\label{non direct-optimal}
    %y_1(\beta;W)&=&-\frac{2\Gamma}{\pi} \int  dk_1  dk_2 \
    %\,\frac {\delta(k_1+k_2-E_u)} {|k_1-Z_x^*||k_2-Z_y|}\nonumber
    y_1&=& \frac{2}{\pi^2} \int    \
    \,\frac{g}  {|k_1+k_2-\beta-ig|^2}\frac{\,dk_1
    dk_2}{|k_1-i||k_2+i|},
    \earray
where $g=\Gamma_u/\Gamma$. Let us study the case of $\beta=0$,
where this function (which is even in $\beta$) achieves its
maximum. Combining this with Eq.~(\ref{gamma},\ref{normalization})
one finds for the optimal $W$
    \be
    \label{optimal gamma}
    \gamma(g;|\Delta|\gg 1,\beta=0;W_{opt})= \half +f(g),
    \ee
which is plotted, as a function of $g$ for perfect color matching,
$\beta=0$, in Fig.~\ref{fig:B_plot} (b). The function $f(g)$ is
negative, monotonically decreasing and vanishes for $g=0$, where
maximal entanglement, $\gamma=\half$, is achieved.  For systems
such as the biexciton radiative cascades, one can not get maximal
entanglement even when the colors perfectly match, since $g\approx
1.5-2$, however, one does get substantial entanglement,
$\gamma\approx 0.4$.

The entanglement $\gamma$ is, of course, sensitive to the matching
of colors across generations, so that when $\beta \gg 1$ the
entanglement becomes small. This is illustrated in
Fig.~\ref{fig:B_plot} (c) which shows $\gamma$ as a function of
$\beta$ for $g=2$, the value relevant to biexciton decay.
%%%%%%%%%%%%%%%%%%%%%%%%%%%%%%%%%%%%%%%%%%%%%%%%%%%%%%%%

%%%%%%%%%%%%%%%%%%%%%%%%%%%%%%%%%%%%%%%%%%%%%%%%%%%%%%%%
%%%%%%%%%%%%%%%%%%%%%%%%%%%%%%%%%%%%%%%%%%%%%%%%%%%%%%%%
%%%%%%%%%%%%%%%%%%%%%%%%%%%%%%%%%%%%%%%%%%%%%%%%%%%%%%%%
Physically, choosing $W$ may be thought of as letting the two
polarizations go through different gates that introduce different,
but fixed, time delays on the two colors. To see this, we note
first that each of the two factors of Eq.~(\ref{optimal}) can
indeed be interpreted as a time shift. This follows from the fact
that
    \be
    \frac{k-Z}{|k-Z|}\approx i e^{-ik/\Gamma} e^{iE/\Gamma}
    \label{eq:approx}
    \ee
for $k \approx E$, see Fig.~\ref{fig:B_plot} (a). This represents
a shift of the wave function in coordinate space by $1/\Gamma$
which can also be interpreted as a (non-random) shift in time by
$1/\Gamma$. Therefore, the two factors in $W_{opt}$ may be
implemented, approximately, by manipulating the optical paths.

It is important to understand that the manipulation of the quantum
chronology proposed here is a fixed unitary manipulation of the wave
function, which is a non-random object. It is not a manipulation of
the individual detection events which are random and uncontrollable.
However, the probability distribution for these random events is
determined by the wave function according to the rules of quantum
mechanics.

\begin{figure}[ht]
\begin{center}
 \vspace{-0.4cm}
 \includegraphics[width=6.8cm]{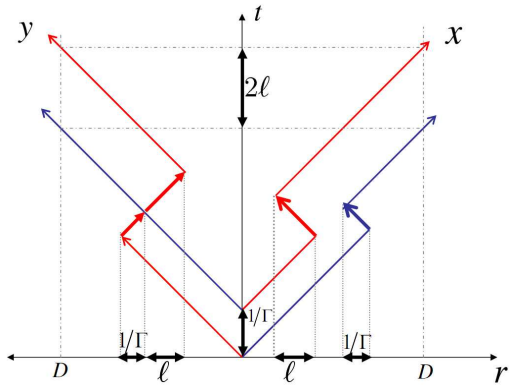}
 \vspace{-0.4cm}
  \caption{(color on line) A space-time diagram representing the path of the photons.
  For clarity, $x$ and $y$ polarized photons
  are drawn as propagating to the left and right, respectively.
  A delay is represented by reflecting a photon back in space.
  An observer located a distance D from the origin cannot
  distinguish between the $x$ and $y$ polarized pairs,
  neither  by their arrival times nor by their energies (for $\beta=0$).
  $\ell$ is an arbitrarily chosen distance, which determines the {\em average} time
  difference between the actual detection of the first
  and second photons.
  }
  \label{fig: space time_plot}
\end{center}
\end{figure}

From Eq.~\eqref{optimal}, we see that $W_{opt}$ has two factors,
affecting the two photons. Fig.~\ref{fig: space time_plot}
explains why both photons must be manipulated: to yield entangled
photons, all the properties distinguishing the $x$ and $y$
polarized photons must be erased. This requires that: the arrival
times at the detector $D$ of photons with energies $E_x$ (the red
photons in Fig.~\ref{fig: space time_plot}) must be independent of
polarization,  and likewise for the photons with energy $E_y$ (the
blue photons in the figure).

Suppose that we let the $y$-polarized photon emitted {\em first}
(red arrow on left side of the the figure), travel a distance
longer by $\ell+ 1/\Gamma$ from the photon emitted {\em second}
(blue on the left) thereby reversing their order ($\ell\geq0$ is
arbitrary). Now the time order (chronology) of both polarizations
agrees, the first photon is blue and the second red. However, the
which path information is not yet erased. For the red photon to
arrive at the same time, the $x$-polarized photon emitted {\em
second} must be delayed by a distance $\ell$. For the blue photon
to arrive at the same time, the $x$-polarized photon emitted {\em
first} must be delayed by a distance $1 / \Gamma$. The average
delay between the time of arrival of the blue and red photon is
then $2\ell$.

%%%%
The extra optical paths can be represented as unitary gates acting
on the photon states by $e^{ikL}$, where $L$ is the path length.
The implementation of the delays described above is given by
    \begin{equation}
   U_x=e^{ik_2/\Gamma}e^{ik_1\ell} \quad
   U_y=e^{ik_1(\ell+1/\Gamma)}.
       \end{equation}
It follows that
%    \begin{equation}
  $  W=U^*_xU_y=e^{ik_1/\Gamma}e^{-ik_2/\Gamma}$%\label{form W}
 %   \end{equation}
which, by Eq.~\eqref{eq:approx}, approximates $W_{opt}$.

%%%%%%%%%%%%%%%%%%%%%%%%%%%%%%%%%%%%
%%%%%%%%%%%%%%%%%%%%%%%%%%%%%%%%%%%%
%%%%%%%%%%%%%%%%%%%%%%%%%%%%%%%%%%%%

From the above discussion, we see that although an exact $W_{opt}$
transformation, Eq.~\eqref{optimal}, may be an experimental
challenge, it should be possible to implement suitable
approximations. A possible optical setting
%involving polarization and
% energy dependent optical paths,
is depicted in Fig.~\ref{fig:
warburton}.

\begin{figure}[t]
\begin{center}
\vspace{0cm}
 \includegraphics[width=6 cm]{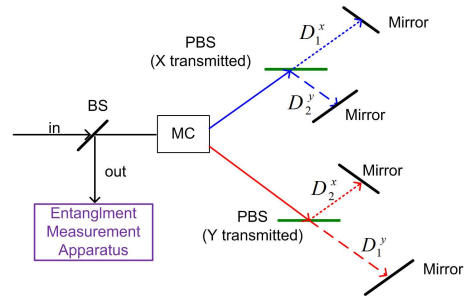}
 %\vspace{-0.2cm}
  \caption{An optical setup which introduces the appropriate delays to each of the photons.
  %The delays are calculated as explained in the discussion leading to Fig.~\ref{fig: space
  %time_plot}, with $L=1/\Gamma$.
  BS (PBS) stands for a (polarizing) beam splitter, and MC for a
  monochromator. The MC (approximately linear) dispersion can be set to
  obtain an approximation to $W_{opt}$ of
  Eq.~(\ref{optimal}). The photons pass the first beam splitter and MC, are reflected back by the
  mirrors and after passing through the MC again, they are measured.
  The optical path are chosen as $2D_1^y=D+1/\Gamma+\ell$, $2D_2^y=D$,
  $2D_1^x=D+1/\Gamma$, and $2D_2^x=D+\ell$ where $D$ is arbitrary.  }
  \label{fig: warburton}
\end{center}
\end{figure}
%\REM{GB}
%One can erase the properties distinguishing between the two
%different channels by letting the photons which were emitted first
%to go an optical path which is longer by $1/\Gamma$ from the path
%the photons emitted second take. In this case, all the photons
%arrive at the same time to the detectors. For such configuration,
%one can choose in Fig.~\ref{fig: warburton} $D_2^x=D_2^y=D$,
%$D_1^x=D_1^y=D+1/\Gamma$, where $D$ is arbitrary. If desired, one
%can let the (say) $E_y$ photons (of condition (a)) arrive earlier
%than the photons with energy $E_x$ (of condition (b)) with a time
%difference $\ell$ by letting the latter go through an additional
%length of $\ell$. For such configuration, choose
%$D_1^y=D_1^y+1/\Gamma+\ell$, $D_2^x=D+\ell$, $D_2^y=D$ and
%$D_1^x=D+1/\Gamma$.
%\REM{EM}
%%%%%%%%%%%%%%%%%%%%%%%%%%%%%%%%%%%%
%%%%%%%%%%%%%%%%%%%%%%%%%%%%%%%%%%%%
%%%%%%%%%%%%%%%%%%%%%%%%%%%%%%%%%%%%

%%%%%%%%%%%%%%%%%%%%%%%%%%%%%%%%%%%%
%%%%%%%%%%%%%%%%%%%%%%%%%%%%%%%%%%%%
%%%%%%%%%%%%%%%%%%%%%%%%%%%%%%%%%%%%
%%%%%%%%%%%%%%%%%%%%%%%%%%%%%%%%%%%%
%%%%%%%%%%%%%%%%%%%%%%%%%%%%%%%%%%%%
%%%%%%%%%%%%%%%%%%%%%%%%%%%%%%%%%%%%

In summary: Entanglement can be created by a non-invasive
(unitary) manipulation of the quantum chronology. This provides a
possible and practical avenue for creating entangled photon pairs
on demand.

\medskip{\bf Acknowledgment:} The research is supported by ISF, by
RBNI, by the fund for promotion of research at the Technion and by
EPSRC (UK). We thank Oded Kenneth and Terry Rudolph for helpful
discussions. D. Gershoni thanks Jonathan Finely for pointing out
his success in tuning colors across generations in biexciton
decays.

%%%%%%%%%%%%%%%%%%%%%%%%%%%%%%%%%%%%

%\bibliographystyle{unsrt}
%\bibliography{bibtex}

\begin{thebibliography}{10}

\bibitem{Nielsen-Chuang}
M.A. Nielsen and I. L. Chuang.
\newblock Quantum Computation and Quanum Information.
\newblock Cambridge U.P., 2000.

\bibitem{down-conversion-Bouwmeester}
K. Mattle {{\em et al}.}
%H. Weinfurter, Dik~Bouwmeester, Jian-Wei~Pan and Anton Zeilinger.
%\newblock Experimental quantum teleportation.
\newblock Nature, 390, 575 (1997).

\bibitem{down-conversion-Kwiat}
P.G. Kwiat, {\em et al.}
%K. Mattle, H. Weinfurter, Anton Zeilinger, Alexander~V. Sergienko, and Yanhua Shih.
%\newblock New high-intensity source of polarization-entangled photon pairs.
\newblock Phys. Rev. Lett., 75, 4337 (1995).

\bibitem{Michler}
P.~Michler, {\em et al}.
%A.~Kiraz, C.~Becher, W.~V.~Schoenfeld, P.~M.~Petroff, Lidong Zhang, E.~Hu, and A.~Imamoglu
%\newblock A Quantum Dot Single-Photon Turnstile Device.
\newblock Science, 290, 2282 (2000).

\bibitem{Santori}
C. Santori, {\em et al.}
%D. Fattal, J. Vukovich , Glenn S. Solomon and Yoshihisa Yamamoto.
%\newblock Indistinguishable photons from a single-photon device.
\newblock {Nature}, 419, 594 (2002).

\bibitem{Yuan}
Z. Yuan, {\em et al}
%B.E.~Kardynal, R.~M. Stevenson, Andrew J.~Shields, Charlene J.~Lobo, Ken Cooper, Neil S.~Beattie, David A.~Ritchie, and Michael Pepper.
%\newblock Electrically Driven Single-Photon Source.
\newblock {Science}, 295,102 (2002).

\bibitem{akopian}
N.~Akopian, {\em et al}.
%N.~H. Lindner, E.~Poem, Y.~Berlatzky, J.~Avron, D.~Gershoni, B.~D. Gerardot, and P.~M. Petroff.
%\newblock Entangled photon pairs from semiconductor quantum dots.
\newblock {Phys. Rev. Lett.}, 96, 130501 (2006).

\bibitem{shields}
R.~J. {Young}, {\em et al}.
%R.~M. {Stevenson}, P.~{Atkinson},  K.~{Cooper}, D.~A. {Ritchie},and A.~J. {Shields}.
%\newblock {Improved fidelity of triggered entangled photons from single quantum dots}.
\newblock {New J. of Phys.}, 8, 29 (2006). R. Hafenbrak
{\em et al}.
%S M Ulrich1, P Michler, L Wang, A Rastelli and O G Schmidt
%\newblock {Triggered polarization-entangled photon pairs from a single quantum dot up to 30K}.
\newblock {\em New J. of Phys.}, 9, 3158 (2007).

\bibitem{Benson}
O. Benson {\em et al}.,
\newblock {Phys.\ Rev.\ Lett.}   84, 2513 (2000).

\bibitem{bayer}
M. Bayer {\em et al}.,
\newblock {Phys. Rev. B} 65, 195315 (2002).

\bibitem{ref:experiments-degeneracy}
B.D. Gerardot, {\em et al}.
%S. Seidl, P.A. Dalgarno Manipulating exciton fine structure in quantum dots with a lateral electric field
\newblock Appl.\ Phys.\ Lett. 90, 041101 (2007).
S. Seidl {\em et al}.
%M. Kroner, A. Hogele Effect of uniaxial stress on excitons in a self-assembled quantum dot
\newblock Appl.\ Phys.\ Lett. 88, 203113 (2006)

%\bibitem{Reime} M. E. Reime et. al., quant-ph/0706.1075

\bibitem{bimberg}
S. Rodt,{\em et al}.
%A. Schliwa, K. Potschke,
%\newblock Correlation of structural
%and few- particle properties of self-organized InAs/GaAs quantum
%dots
\newblock {Phys. Rev. B}, 71, 155325 (2005)

\bibitem{finley}
Jonathan J. Finley, private communication

\bibitem{Reimer}
M. E. Reimer {\em et al}., quant-ph/0706.1075


\bibitem{Cohen-Tannoudji}
C.~Cohen-Tannoudji, J.~Dupont-Roc, and G.~Grynberg.
\newblock {Atom-Photon Interactions}.
\newblock John Wiley \& Sons, 1992.

\bibitem{horodecki-review}
R. Horodecki, {\em et al}.
%\newblock Quantum entanglement,
quant-ph/0702225.

\bibitem{peres_test}
A. Peres.
%\newblock Separability criterion for density matrices.
\newblock {Phys. Rev. Lett.}, 77, 1413 (1996).

\bibitem{entanglement_measure}
G.~Vidal, R.~F. Werner
%\newblock Computable measure of entanglement.
\newblock {Phys. Rev. A}, 65, 032314 (2002).

\bibitem{meirom}
E.~A. Meirom, {\em et al}.
%N.~H. Lindner, Y.~Berlatzky, E.~Poem, N.~Akopian, J.~E. Avron, and D.~Gershoni.
%\newblock Distilling entanglement from cascades with partial
%``which path'' ambiguity.
arxiv.org:0707.1511 (2007).

\end{thebibliography}

%\begin{thebibliography}{10}
%\end{thebibliography}

\end{document}